\documentstyle[epsfig]{jaa}
\begin{document}
\title [Associated absorption in B2 0902+343]
{Associated HI absorption in the z=3.4 radio galaxy B2 0902+343
observed with the GMRT}

\author[Chandra et al.]
{Poonam Chandra$^{1,2}$\thanks{e-mail:poonam@tifr.res.in}, 
Govind Swarup$^3$\thanks{e-mail:swarup@ncra.tifr.res.in},
\newauthor
Vasant K. Kulkarni$^3$\thanks{e-mail:vasant@ncra.tifr.res.in}, \&
Nimisha G. Kantharia$^3$\thanks{e-mail:ngk@ncra.tifr.res.in}\\
$^1$ Tata Institute of Fundamental Research, Mumbai 400 005, India\\
$^2$ Joint Astronomy Programme, IISc, Bangalore 560 012, India\\
$^3$ National Centre for Radio Astrophysics, TIFR, Pune 411 007, India}

\maketitle

\begin{abstract}
We have made observations of the associated HI absorption of a high redshift
radio  galaxy 0902+34 at z=3.395 
with the Giant Meterwave Radio Telescope in the $323\pm1$
MHz band. We find a narrow
absorption line with a flux density of 11.5 mJy at a  redshift of 3.397
consistent with that observed by Uson et al. (1991), Briggs et al. (1993)
and de Bruyn (1996).
A weak broad absorption feature reported by de Bruyn (1996) has
not been detected in our observations. We also place an upper limit
of 4 mJy (2 $\sigma$) on
 emission line strength at the position where
Uson et al. (1991) claimed to have found a narrow emission line.
\end{abstract}
                                                                                
                                                                                
\section{Introduction}

It is believed that the epoch of galaxy  formation lies beyond z $\geq$5
but there is considerable uncertainty in their subsequent evolution.
  Radio galaxies and quasars at high redshifts are useful in studying
the early stage of formation of galaxies. Many high redshifted
radio galaxies and quasars have been looked for 
associated absorption of continuum radiation by neutral hydrogen (HI).
Associated absorption provides invaluable information about the
environment surrounding
astronomical objects. This can be studied
either by redshifted 21-cm absorption of HI against the radio continuum
or by narrow absorption in the spectra of Ly-$\alpha$  emission
or absorption. Till
date there is only one object beyond redshift z $\geq$3 in which associated
21-cm absorption has been found.
This is the
radio galaxy B2 0902+343 at a redshift of z=3.395 in which associated
absorption was found by Uson et al. (1991) and later confirmed by
Briggs et al. (1993) and de Bruyn (1996).

The radio galaxy B2 0902+343 was first identified with an optical
object  by Lilly(1988)  under
``1 Jansky empty field survey''.
Its very flat optical broad spectral energy distribution indicates
that it may be a 'protogalaxy', undergoing its first major burst of star
formation (Einsenhardt \& Dickinson 1992).
The spatial extension of this radio source is $7"\times 4" $.
The galaxy shows a bizarre radio structure in high resolution
radio images (Carilli 1995). The radio source has  two outer lobes 
and a prominent nuclear core source
with a flat spectral
index. The flux density of the core at 1.665 GHz is $\sim 10$ mJy, being about
2.4 \% of the strength of the entire source at that frequency.
At the northern end of a sharp knotty jet emerging from the core,
with a sharp 90$^o$ 
bend towards its outer end, 
is located a prominent hot spot and also a long 'plume' of emission
with an ultra steep spectral index of -3.3. The southern
lobe consists of two 
components having perpendicular orientation with respect to the 
direction of the core. The source
seems to be oriented at an angle of about  45$^o$
to 60$^o$ with respect to
the sky plane (Carilli 1995). Moreover this is one of the unique
high redshifted sources which does not follow radio-optical alignment effect.
HST observations by Pentericci et al. (1999) show that 
the optical galaxy consists
of two regions of approximately the same flux density with a void 
or valley in between.
Astrometry measurements show that the radio core is situated in the valley
(Pentericci et al. 1999) but do not rule out the possibility
that the radio core could be coincident
with any of the two components.
The presence of narrow Ly-$\alpha$ emission 
distinguishes it from a high redshifted
quasar. Also, recent Chandra observations by Fabian et al. (2002) 
has found
X-ray emission from the source. The centroid of the X-ray emission
(J2000 09 05 30.17, +34 07 56) is coincident with the
flat spectrum nuclear core source (J2000 09 05 30.13, +34 07 56.1),
which suggests it to be an AGN.

This peculiar galaxy and its surroundings have been searched for absorption
and emission features by several workers using
various radio telescopes. Uson et al. (1991)
 detected a narrow absorption line of
11 mJy associated with B2 0902+343. In addition, they claimed to have
found an emission line of 11 mJy located about 33' away from the galaxy,
which was
ascribed to emission by a massive HI condensate as
expected for a Zeldov'ich 'pancake'.
 Subsequent observations by Briggs et al. (1993) using the 
Arecibo Radio Telescope and by de Bruyn (1996) using Westerbork 
Radio Telescope (WSRT) confirmed the presence of the narrow
associated line but did not detect the emission line.
The possible presence of a broad
absorption feature extending to several hundred km s$^{-1}$ on the blue-ward
side of the narrow absorption feature has been reported by de Bruyn (1996).
 Cody et al. (2003) looked for the presence of OH gas in the galaxy to
get an insight in the molecular content of the galaxy but did not 
detect redshifted 1665/1667 MHz OH lines. 
They put a 1.5 $\sigma$ upper limit of 3.6 mJy, which gives an
upper limit on the OH column density of $N(OH)\le 10^{15}$ cm$^{-2}$.

We have observed the radio galaxy B2 0902+343 with the Giant Meterwave Radio
Telescope (GMRT) in the band $323\pm1$ MHz for studying the
associated absorption features.
We have detected the narrow absorption 
line component. However, the broad absorption
feature claimed by de Bruyn (1996) is absent in our results. 
We also place a 2 $\sigma$ upper limit of 4 mJy at the position of
the emission line claimed by Uson et al. (1991). We present observations and
data analysis in Section 2. Results are given in Section 3 and
discussions and conclusions in Section 4.
	
\section{Observations and data analysis}

Observations of the radio galaxy B2 0902+343 were made with
the GMRT in March 1999 and  on May 15, 2000 near 323 MHz. GMRT is
an aperture synthesis radio telescope situated about 80 km north of
Pune in India. It consists of 30 fully steerable parabolic dishes of 45
meter diameter each, spread over an extent of about 25 km
(Swarup et al. 1991). Fourteen out of the
30 antennas are located within a compact array of about 1 square km
in size and the remaining
16 antennas are located along a Y-shaped array, with each arm of about
14 km length.

In March 1999, observations of the radio galaxy B2 0902+343
were made for $\sim$ 8 hours with
12 to 15 antennas. Although observations were made with 128 channels
over  a bandwidth of 2 MHz centered at 323 MHz, the data was good 
for  only the central part of 1 MHz. 3C286 was used as flux
calibrator and 3C216 and 0834+555 were used as phase calibrators.
RMS noise of the continuum map after collapsing 40 channels was 4 mJy.
RMS noise on the line images, after subtracting the continuum was 1.4 mJy.

On  May 15 2000, observations were made for 6 hours on the source, when only
15 good antennas were available. The observations were
made with 128 channels over a bandwidth of 2 MHz.
Therefore, the width of each channel was 15.6 kHz.
3C48 and 3C286 were used as flux calibrators. 3C216 was used as a
phase calibrator. 3C48, 3C286 and 3C216 were also used as bandpass
calibrators. The phase calibrator was observed for 8 minutes after 
every 40 minutes.

The data was analyzed using Astronomical Image Processing System (AIPS)
of NRAO.
The  gains of the antennas were determined using the flux and phase calibrators 
after flagging any bad data. The flux densities of 
3C48 and 3C286 was calculated
to be 43.64 Jy and 26.03 Jy respectively, on the scales of Baars et al. (1977)
using the AIPS task SETJY. The flux density of 3C216 was found to be 
$16.65 \pm 0.21$ Jy. After bandpass calibration using the flux and 
phase calibrators, AIPS task SPFLG was run on the source 0902+34 to 
flag any data showing RFI. In total, the amount of good
data that could be used finally was 76\%.  We then averaged several line free 
channels and made a continuum map of the source using AIPS task 
IMAGR. We also did self calibration in order to get rid
of residual phase errors and to improve the 
quality of the map.  
The resolution of the map was $39" \times 30"$. The continuum flux  
density of the source was found to be $1.357\pm 0.008$ Jy. The dynamic range
obtained in continuum map  was $\sim 450$. The source
is not resolved in our map. The continuum map was then subtracted 
from the line data using UVSUB. AIPS task UVLIN was then run and the line free
channels were used for the baseline fitting. The UVLIN output was 
used for making a spectral cube of the data. The narrow absorption 
feature towards B2 0902+343 was clearly detected but no other
absorption or emission feature was seen in the spectral
cube covering the field of view of the 45-m antennas of the GMRT.

\begin{figure}
\epsfig{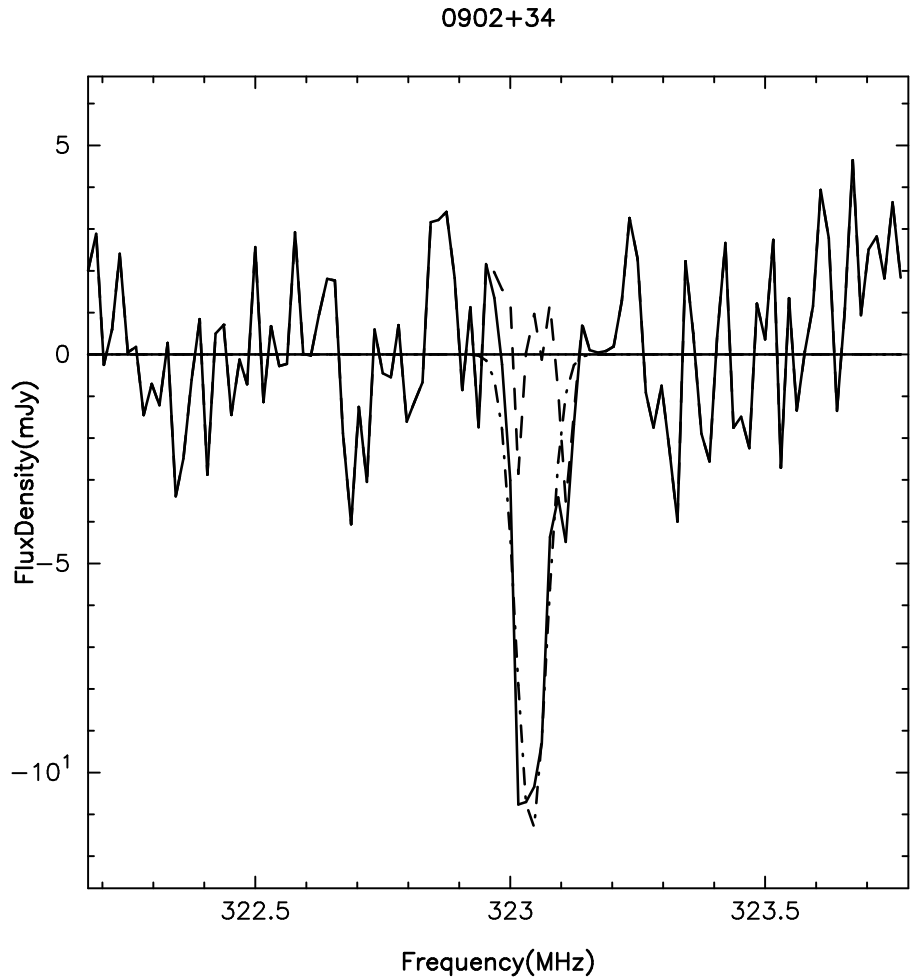}
\caption{One-component Gaussian fit to the absorption line observed 
on 15th May,
  2000. Dot-dashed line shows gaussian fit to the observed absorption feature.
Dashed line show 
residual to the Gaussian fit.}
\label{fig:gauss_pc}
\end{figure}

\section{Results}

\subsection{HI Absorption Narrow Line}
Fig. \ref{fig:gauss_pc}  shows the HI absorption profile obtained from  
the observations made on 15th May 2000. 
The dashed dot line shows a one-component Gaussian fit
to the observed absorption line. 
We find that there is a continuum
offset of about 1 mJy.  We have made  one-component Gaussian fit
after removing this "continuum" of 1 mJy offset and 
the results are summarized in Table \ref{tab:gauss}.  A
possible weak absorption feature reported by Briggs et al. (1993) near 322.5
MHz is not seen in our observations.

\begin{figure}
\centering
\epsfig{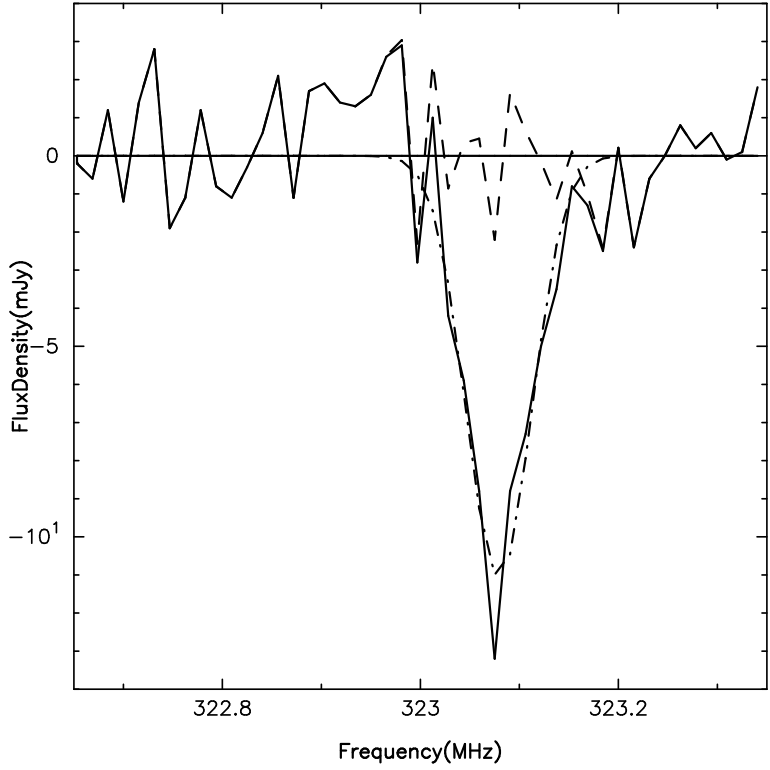}
\caption{One-component Gaussian fit to the absorption
line observed on March,1999. Dot-dashed line shows a fit to 
the observed absorption feature and dashed line shows
 residual to the Gaussian fit.}
\label{fig:gauss_ngk}
\end{figure}

The absorption profile for the March 1999
observations
is shown in Fig. \ref{fig:gauss_ngk}. This spectrum
shows the presence of a small asymmetric
profile towards the higher frequency.  Hence, we have made both
one-component and two-component Gaussian fit to this data.
In table 1, we summarize the results of only 1-component
gaussian fit, since we find from 
the two-component 
gaussian fit that the asymmetry towards higher frequency end of the 
abosrption line is not significant.

\begin{table}
\scriptsize
\vspace*{1cm}
\caption{Best fit Gaussian parameters for the  March,1999 and  May,2000 data}
\label{tab:gauss}
\begin{tabular}{ccccccc}
\hline
\hline
Date of & Gaussian & Line Depth & position of line & redshift & FWHM & rms\\
Observn & fit & mJy &            MHz &   z        & kms$^{-1}$ & mJy/beam\\

\hline
2000 May& 1-comp. & 11.5 & 323.062$\pm$0.005 & 3.397 & 76$\pm$6 & 1.8\\
1999 Mar & 1-comp. & 11.1 & 323.058$\pm$0.005 & 3.397 & 72$\pm$6 & 1.4\\
\hline
\end{tabular}
\end{table}

\begin{figure}
\centering
\epsfig{file=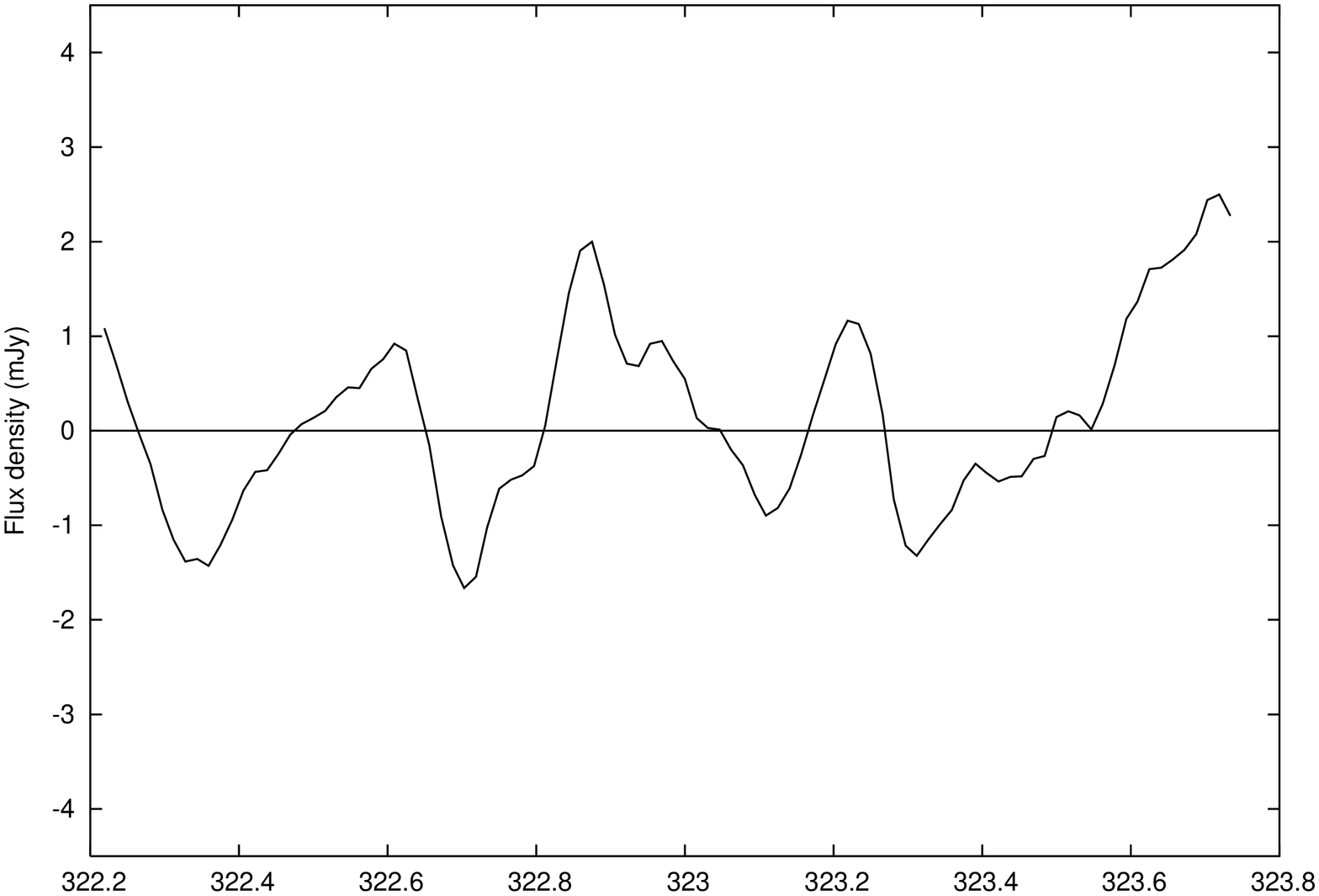, height= 9cm, angle=-90}
\epsfig{file=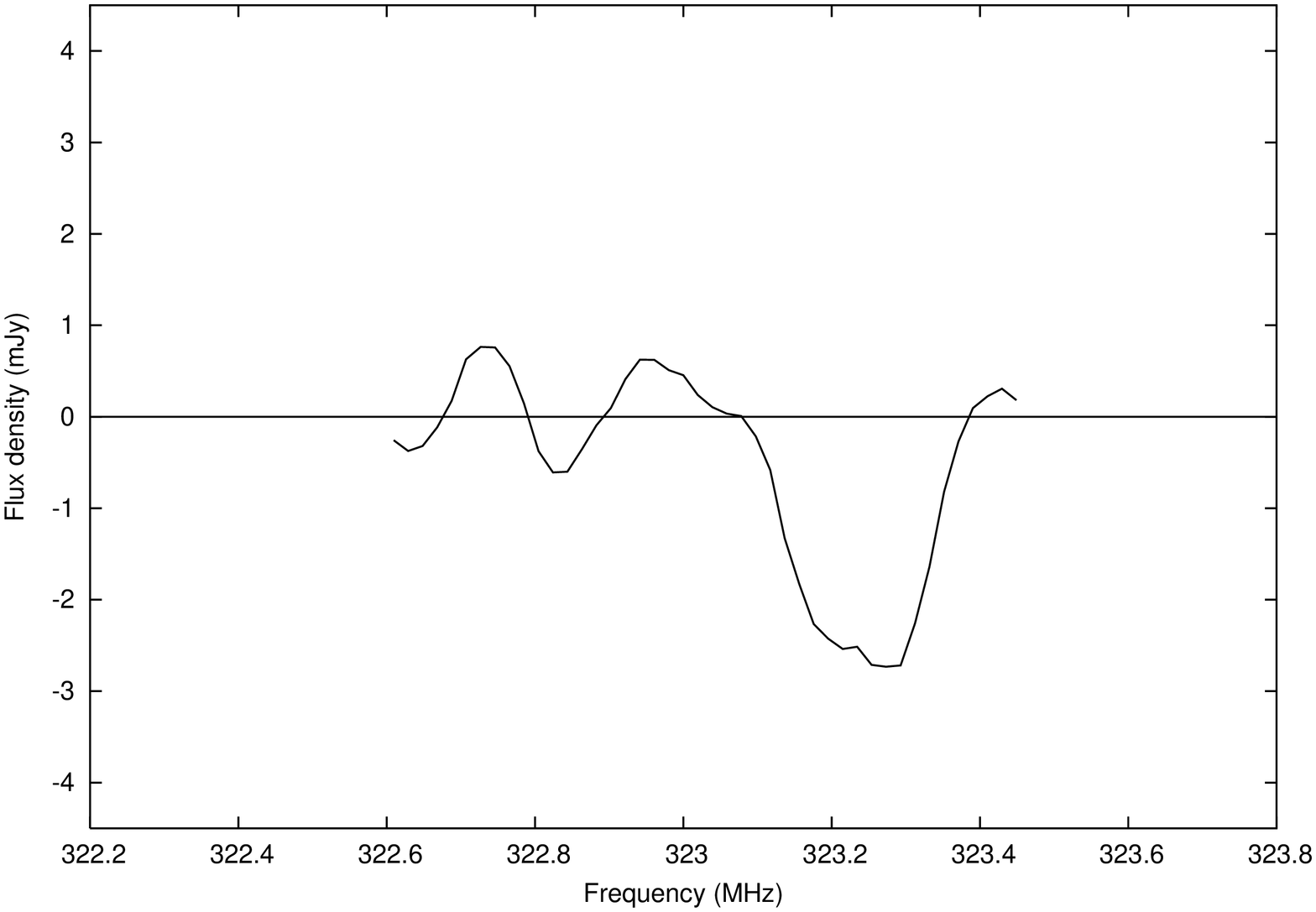, height= 9cm, angle=-90}
\caption{Upper panel shows the residuals of one-component Gaussian fit 
to absorption line observed on May, 2000. 
Lower panel shows the same for de Bruyn's dataset.}
\label{fig:residual}
\end{figure}

\subsection{Broad HI Absorption Line}

de Bruyn (1996) reported possible presence of a weak broad absorption
feature extending several hundred km s$^{-1}$ blue-ward (
at a higher frequency) of the
prominent narrow absorption line.  In order to search for its presence in
the GMRT observations, we have removed the Gaussian fit shown in Figs. 
\ref{fig:gauss_pc} and \ref{fig:gauss_ngk}
 and then convolved the residuals with a Gaussian profile of
FWHM of 100 km s$^{-1}$.  
The smoothed residual spectra of the GMRT observations 
on May 15, 2000 are shown in Fig
\ref{fig:residual}.
No broad absorption feature is visible in either spectra.
Since a digital file of the spectrum observed
with WSRT by de Bruyn was not saved by him (private communication), we
have carefully digitized the data from the spectrum published by him as
Figure-1 (de Bruyn 1996).  We have fitted a Gaussian to the
narrow feature of his spectrum and found that the narrow absorption line
has a FWHM of $80\pm8$ km s$^{-1}$. We have also convolved the
residuals of his data 
with 100 km s$^{-1}$ FWHM as shown in the last panel of 
Fig. \ref{fig:residual}.  Fig. \ref{fig:residual} shows
possible presence of a broad absorption feature at 3
$\sigma$ level in the WSRT
observations.  However, we place an upper limit of 3 mJy at 3
$\sigma$ level from the GMRT
data of 15 May 2000.

\subsection{Non-detection of HI Emission}

\begin{figure}
\epsfig{file=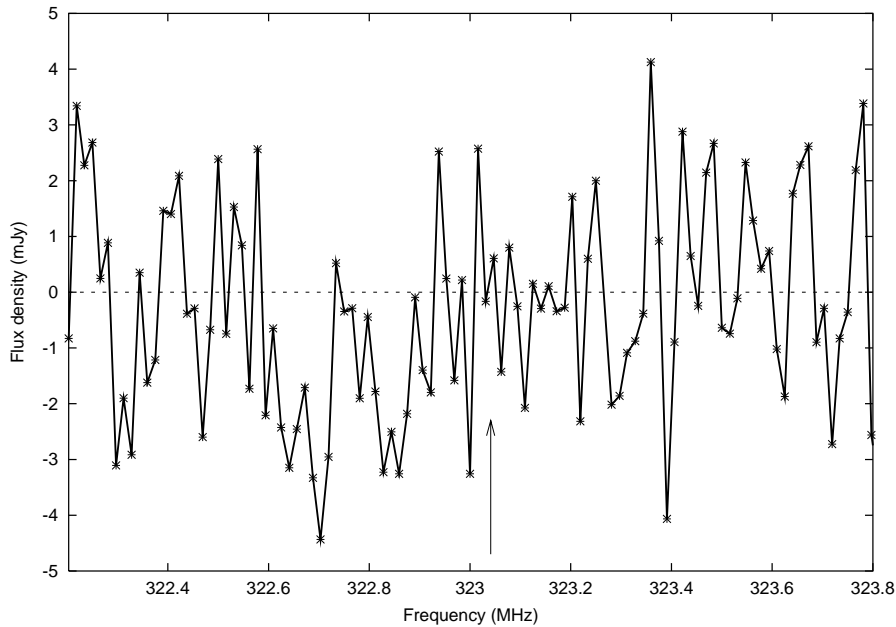, height= 12cm, angle=-90}
\caption{Spectrum at the position of emission  claimed by Uson et al. (1991)
at 09$^h$03.8$^m$, 33$^0$ 52' (B 1950) and frequency of about 323.0 MHz. 
Arrow point towards the frequency at which Uson et al. found the emission.}
\label{fig:emission}
\end{figure}

Uson et al. (1991) had claimed the detection of an HI
emission feature of 11 mJy at a frequency of 323.041 MHz, towards the
direction of $\alpha= 09^h 03.8^m,\, \delta= 33^o 52'$
 (B 1950), which is 33' away from the
position of the source B2 0923+343.  We made the spectral cube with 200"
resolution (same resolution as used by Uson et al. (1991))
and corrected it for primary beam pattern. We do not
see any emission feature at the given position. We place a 2$\sigma$ upper
limit of 4 mJy at the position of the above emission feature. The GMRT
spectrum towards this direction is shown in Fig. \ref{fig:emission}. 

\subsection{Search for HI absorption towards Background Sources}

Although the probability of a strong radio source being located in a
cluster near B2 0902+343 is small, we have searched for any absorption
features towards 3 radio sources in the field of view of the primary beam
of the 45 m dishes of the
GMRT (See Table \ref{tab:other} ). No absorption lines were detected.

\begin{table}
\centering
\caption{The sources in the FoV, looked for any possible absorption
  feature}
\label{tab:other}
\begin{tabular}{ccc}
\hline
\hline
Source & Position & Flux density\\
  & J2000 & Jy\\
\hline

1 & 09 07 17.6, +33 44 10 & 0.675$\pm$0.008\\
2 & 09 06 19.4, +34 17 32 & 0.289$\pm$0.008\\
3 & 09 04 32.7, +34 32 12 & 0.183$\pm$0.008\\
\hline
\end{tabular}
\end{table}

\section{Discussions and Conclusions}

Our observations show that the depth of the HI absorption 
at a redshift (doppler shift corrected)
 of $3.3967\pm0.0002$ is 11.5 mJy with a FWHM of
$74\pm6$ km s$^{-1}$ (see Tables \ref{tab:gauss}  and \ref{tab:all}).
In our GMRT observations with only 15 antennas, we could not
get sufficient resolution to locate the position of the line.
The possible scenarios for its position could be either the northern
 hot spot or a tori/jet surrounding the nuclear
source. The observed depth of the narrow absorption line
of 11.5 mJy is close to the observed flux
density of the radio nucleus (flux density of 9.5 mJy at
1.65 GHz), implying 100\% absorption (optical depth of 1) by 
cold HI gas which could be present in a disk or tori.
Associated HI absorption has been found in many compact steep spectrum
sources (Vermeulen et al. 2003). However, only absorptions  upto about
10-40 \% have been seen in these sources. 
Assuming that the surrounding disk has a scale size such that it covers 
a possible steep spectrum core and part of the base of the jet, 
Rottgering et al. (1999) have suggested that the observed absorption
may be associated with a steep spectrum core. In that case the
required optical depth could be much less, say 20\%.

Although associated HI absorption is generally not seen in extended 
powerful radio galaxies, there are some radio galaxies in which 
HI absorption is seen against the hot spots (Morganti 2002 and references 
there in).   
de Bruyn (1996) has suggested
that the absorption is likely to be against the northern hot spot.
The hot spot has an angular size of 0.2" and contains about 75\% of
the total flux density. Its extrapolated  
flux density at 323 MHz is $\sim$ 1 Jy. In this case optical
depth of only 1.2\% is required, which is pretty reasonable.
 The implied column 
density, N(HI) $\sim 1.6 \times 10^{21}$ cm$^{-2}$, assuming 
optical depth of 0.012 and a mean spin temperature of 1000 K.
The HI gas could be surrounding the hot  
spot which is not an unlikely scenario considering that the optical
observations of B2 0902+343 indicates it to be a protogalaxy.
The source may be associated with a 
merging galaxy located  
near the hot spot or dwarf galaxy along the line of sight,
as a possible scenario suggested by Rottgering et al. (1999).


Considering the above discussion, it seems to us that the HI absorption
could be associated with the hot spot. However, sub arcsec
high-resolution observations
 at low frequencies are required  in order to determine
the location of the absorber.

\begin{table} 
\footnotesize 
\caption{Details of all observations of 0902+34 by
  various workers}
\label{tab:all}
\begin{tabular}{ccccccc}
\hline
\hline
Obs. Date & Telescope & time & Line depth & FWHM . & rms & References\\
          &             & Hours & mJy & kms$^{-1}$& mJy/beam\\
\hline
Apr,91 & VLA & 14 &  11.4& 270$\pm$50  & 1.8& Uson et al. 1991\\
1992   & Arecibo &  & 14 & 90 &  1.9 & Briggs et al. 1993\\
1995 & WSRT & 144$^{*}$ &   12 & 100$^{**}$ & 1 & de Bruyn 1996\\
May,00 & WSRT & 12 & 15.9 &120$\pm$50 & 2.4& Cody et al. 2003\\
Mar,99 & GMRT & 8 & 11.1 & 72$\pm$6 & 1.4 & this paper\\
May,00 & GMRT & 6 & 11.5 & 76$\pm$6 & 1.8 & this paper\\
\hline
\end{tabular}

$^{(*)}$ As mentioned in Rottgering at al. (1999).\\ 
$^{(**)}$ Our estimate to the data of de Bruyn data gives line depth
of $9.93 \pm 1.45$ mJy,  FWHM
of $80\pm 8$ km s$^{-1}$.

\end{table}
We conclude with the following results:
\begin{itemize}
\item { We have detected a narrow absorption line with the
GMRT in B2 0902+343
with flux density of 11.5 mJy which  has a half-power velocity width of
74$\pm$6 km s$^{-1}$. The  line depth is consistent with previous
  observations.}

\item{ The broad absorption feature indicated by de Bruyn (1996) is
 not seen in our observations. We also do not detect the emission
 line (33' away) as  claimed by Uson et al. (1991) which is
consistent with the results
  by Briggs et al. (1993) using the Arecibo Radio telescope
and de Bruyn (1996) using the WSRT.}
\end{itemize}

\section*{Acknowledgments}
We thank the staff of the GMRT that made these observations possible.
GMRT is run by the National Centre for Radio Astrophysics of the 
Tata Institute of Fundamental Research.
We acknowledge  valuable comments by Prof. Rajaram Nityananda.
N. G. Kantharia thanks late Prof. K. R. Anantharamaiah for suggesting
the observation of this source in 1999. We thank the anonymous
referee for his/her useful comments.

\end{document}